\documentclass{article}%
\usepackage{amssymb}
\usepackage{amsmath}
\usepackage{amsfonts}
\usepackage{graphicx}%
\providecommand{\U}[1]{\protect\rule{.1in}{.1in}}

\begin{document}

\title{On tunnelling through the black hole horizon}
\author{V. A. Belinski\\\textit{ICRANET, 65122 Pescara, Italy;}\\\textit{Phys. Dept., Rome University "La Sapienza", 00185 Rome, Italy; \ \ \ }\\\textit{IHES, F-91440 Bures-sur-Yvette, France.}\\\textit{E-mail address: belinski@icra.it}}
\date{}
\maketitle

\begin{abstract}
It is shown here that there is no way for particle creation to occur by
quantum tunneling through an infinitesimal neighborhood of the black hole
horizon. This result is a trivial consequence of the regularity of the
horizon, the equivalence principle and the general covariance of the
relativistic theory of gravity. Moreover, we also confirm the less trivial
statement that no particle creation by quantum tunneling through the black
hole horizon is possible independent of the size of the presupposed tunneling domain.

\end{abstract}

\section{ On some delusions existing in the literature}

During the last twelve years a series of papers has appeared trying to
introduce particle creation by the Schwarzschild black hole as a
quasiclassical tunneling process through an \textit{infinitesimal
neighborhood} of the black hole horizon at the radius $r=r_{g}$, i.e., through
the region between $r_{g}-0$ and $r_{g}+0$. The first papers containing such
an idea were \cite{V} and \cite{PW}\footnote{The paper \cite{PW} has the
pretension to do something essential with the back reaction of the particles
created by the tunneling in the black hole spacetime. Moreover, the author's
last statement claims that without back reaction, i.e., in the fixed external
gravitational field, the tunneling \textquotedblleft would appear to be
precluded". In fact this does not follow from their result. If it were so, the
imaginary part of the action they obtained would disappear when the back
reaction vanishes. Instead, in such a limit their final formula gives the
standard result. Then at best this paper can be considered as an attempt to
calculate the back reaction corrections to the conventional fixed external
field approach. In the first approximation there is no difference between
\cite{PW} and all other papers of this sort.} and these gave birth to many
successors among which are \cite{ANVZ, NVZ, CHNVZ}. The arguments of all these
papers proceed as follows. In the Schwarzschild black hole space-time covered
by the Schwarzschild coordinates (which we take only for definiteness, the
results of our analysis in no way depend on the choice of coordinates, see
below) the metric is:%
\begin{equation}
-ds^{2}=-\left(  1-\frac{r_{g}}{r}\right)  c^{2}dt^{2}+\left(  1-\frac{r_{g}%
}{r}\right)  ^{-1}dr^{2}\,, \label{1}%
\end{equation}
then the particle action has the form%
\begin{equation}
S=-Et+W(r)\,, \label{2}%
\end{equation}
and Hamilton-Jacobi equation for the function $W(r)$ corresponding to the
\textit{outgoing} trajectories gives:
\begin{equation}
\frac{dW}{dr}=\frac{\sqrt{E^{2}r^{2}-m^{2}c^{4}r(r-r_{g})}}{c(r-r_{g})},\text{
\ \ }E\geqslant mc^{2}\,, \label{3}%
\end{equation}
where the square root takes positive values. The assertion is that any
outgoing trajectory, corresponding to some fixed value of the arbitrary
constant $E$, consists of two pieces: one is located in the external space
where $r>r_{g}$ and the other inside the horizon where $r<r_{g}$ but
\textit{each of these two pieces starts at some point of the horizon }%
$r=r_{g}$. Classically a particle cannot propagate between the inside and
outside of the horizon along such a trajectory because of the pole singularity
in the momentum $p(r)=dW/dr$ at points on the horizon at $r=r_{g}$ as can be
seen from (\ref{3}). However, quantum mechanically the jump is possible and
the quasiclassical path connecting both pieces of the outgoing trajectory goes
from $r_{g}-0$ to $r_{g}+0$ through the complex plane of the coordinate $r$
along the infinitesimal semicircle around the singular point $r=r_{g}$. As a
result the function $W(r)$ acquires an imaginary additive term which is
interpreted as the imaginary part of the action $S$ responsible for the
tunneling from inside to outside the horizon.

It is easy to see that this proposal is erroneous. First of all it is
necessary to specify which horizon we are talking about, the future or the
past, because the equation $r=r_{g}$ corresponds to both cases. If we keep in
mind a black hole formed by collapse then only the \textit{future horizon} can
be of relevance. However, it is well known that no outgoing geodesic can have
an external part starting at a point of future horizon and no outgoing
geodesic can have an internal part which makes contact with this horizon from
the inside. This is simple consequence of the geodesic equations. For the
function $W(r)$ in the vicinity of points at $r=r_{g}$ we have:%
\begin{equation}
W(r)=Ec^{-1}r_{g}\ln\frac{r-r_{g}}{r_{g}}\,,\text{ \ \ }r-r_{g}\rightarrow0\,.
\label{4}%
\end{equation}
Since the equations of motion follow from the relation $dS/dE=const.,$ the
equation for trajectories near the horizon is:%
\begin{equation}
ct=r_{g}\ln\frac{r-r_{g}}{r_{g}}+const.\,,\text{ \ \ }r-r_{g}\rightarrow0\,,
\label{5}%
\end{equation}
from which it is obvious that at $r\rightarrow r_{g}$ \textit{the
Schwarzschild time }$t$\textit{\ along any external trajectory tends to minus
infinity while the future horizon corresponds to plus infinity}. Therefore the
classical pieces of any outgoing trajectory have no chance to touch the future
horizon, namely there is no way for them to meet in the infinitesimal region
between $r_{g}-0$ and $r_{g}+0$. Thus in the physical case of a black hole
formed by collapse, the tunneling of the type proposed in \cite{V}%
--\cite{CHNVZ} can never happen.

It would be different story if equation $r=r_{g}$ we understood as the
equation of the \textit{past horizon }of an eternal black hole (ignoring the
fact that such an extension of the Schwarzschild metric has no relation to
physical collapsing objects). In this case the outgoing trajectories indeed
cross the set of points where $r=r_{g}$, namely the set where in accordance
with Eq.~(\ref{5}) the Schwarzschild time tends to minus infinity. However,
this singularity has nothing to do with any tunneling because the outgoing
trajectories crossing the past horizon correspond to particles freely
propagating in a purely classical way from the \textquotedblleft white hole"
region of an eternal black hole to its external space without any obstacles at
the points where $r=r_{g}$. The pole singularity in the function $dW/dr$ at
these points represents a purely coordinate effect without any physical
meaning. In the vicinity of these points, as is obvious from the Kruskal
representation of the extended Schwarzschild metric, the behavior of the
outgoing trajectories is the same as of trajectories of free particles in flat
space-time. Consequently the change in the action along an outgoing trajectory
crossing the past horizon must be real, namely the pole in the integrand of
the function $W(r)$ cannot generate any imaginary part in the action function
$S(r)$. That this is indeed the case can be seen by substitution of the
asymptotic expression for the variable $t$ from (\ref{5}) into the formula
(\ref{2}) for the action. Taking into account the asymptotic behavior
(\ref{4}) of $W(r)$ in the vicinity of the horizon we see that in the action
$S$ the logarithmic terms, originating from the pole in (\ref{3}), cancel each
other and all the remaining terms in the action are regular and real as they
should be.

In other words, the reality of the action in the case under consideration is
due to the fact that in the black hole space-time the Schwarzschild variable
$t$ (we stress that it is just a \textit{variable} since \textit{globally}
\textit{it is not a time function}) acquires an imaginary additive term in the
region inside the horizon (which can be seen from the transformation between
Schwarzschild and Kruskal coordinates). If one considers a particle traveling
between the inside and outside of the horizon, then this imaginary part coming
from the variable $t$ should also be taken into account in the action
(\ref{2}). The foregoing argument shows that it cancels exactly the imaginary
part which is coming from the function $W(r)$.\footnote{The significance of
this fact in the context of the proposals of the articles \cite{V}%
--\cite{CHNVZ} has been pointed out in \cite{B1} and analyzed in more detail
in \cite{P1,B2}.
\par
There were objections to the articles \cite{P1,B2} from the authors of
\cite{CHNVZ}. The response to the objections to \cite{P1} has been given in
\cite{P2,P3}. All pertinent answers to the objections to \cite{B2} expressed
in \cite{Hayw} can be found partly in \cite{P2,P3} and partly are contained in
the text of the present article.
\par
It is necessary to stress that the criticisms expressed in \cite{Hayw} are
groundless and in no way can change any of the statements of paper \cite{B2}
and of the present article.}

The same results emerge in Eddington-Finkelstein, Painleve-Gullstrand or any
other coordinates. One can easily see that \textit{independent of the choice
of coordinates, no part of any outgoing trajectory can make contact with the
future horizon, and the pole singularity in the particle's momentum along such
geodesic can arise only at the points of the past horizon where it represents
a trivial coordinate effect which has no physical significance.} Some
pedagogical attention is deserved by the Painleve-Gullstrand coordinate system
since a widespread delusion can be found in the literature (primarily in
articles \cite{V} and \cite{PW}) that for the problem under consideration the
Painleve-Gullstrand time coordinate is fundamentally better then the
Schwarzschild one. In fact any coordinates can be used (if used correctly, of
course, namely properly extended and properly interpreted) and it makes no
sense to hope one can obtain different physical results just passing to
different coordinates. The Painleve-Gullstrand radial coordinate is the same
as the Schwarzschild $r$ and the Painleve-Gullstrand time coordinate $t_{p}$
can be introduced by the relation:
\begin{equation}
ct_{p}=ct+2\sqrt{r_{g}r}+r_{g}\ln\frac{\sqrt{r}-\sqrt{r_{g}}}{\sqrt{r}%
+\sqrt{r_{g}}}\,. \label{6}%
\end{equation}
The Painleve-Gullstrand form of the Schwarzschild solution is:%
\begin{equation}
-ds^{2}=-\left(  1-\frac{r_{g}}{r}\right)  c^{2}dt_{p}^{2}+\frac{2\sqrt{r_{g}%
}}{\sqrt{r}}cdt_{p}dr+dr^{2}\,. \label{7}%
\end{equation}
The action in these new variables can still be written as $S=-E_{p}t_{p}%
+W_{p}(r)$, where $E_{p}=const.$ and particle's momentum is $p_{p}%
(r)=dW_{p}/dr$. Again we are interested in the \textit{outgoing} trajectories
for which $E_{p}\geqslant mc^{2}$ and in the region $r>r_{g}$ with momentum
$p_{p}(r)>0$. For such trajectories the Hamilton-Jacobi equation written in
the metric (\ref{7}) gives:%
\begin{equation}
\frac{dW_{p}}{dr}=\frac{E_{p}\sqrt{rr_{g}}}{c\left(  r-r_{g}\right)  }\left(
1+\sqrt{1+\frac{\left(  E_{p}^{2}-m^{2}c^{4}\right)  \left(  r-r_{g}\right)
}{E_{p}^{2}r_{g}}}\right)  \,, \label{8}%
\end{equation}
where both square roots are positive. Near the horizon $r=r_{g}$ we have the
asymptotic expression:%
\begin{equation}
W_{p}(r)=2E_{p}c^{-1}r_{g}\ln\frac{r-r_{g}}{r_{g}}\,,\text{ \ \ }%
r-r_{g}\rightarrow0\,. \label{9}%
\end{equation}
The equation of the trajectories follows from the relation $-t_{p}%
+dW_{p}/dE_{p}=const.,$ which in the vicinity of the horizon gives \
\begin{equation}
ct_{p}=2r_{g}\ln\frac{r-r_{g}}{r_{g}}+const.\,,\text{ \ }r-r_{g}%
\rightarrow0\,. \label{10}%
\end{equation}
This equation demonstrates the same result: if the particle trajectory
approaches the horizon from the outside ($r\rightarrow r_{g}$ and $r>r_{g}$)
the Painleve-Gullstrand time goes to minus infinity and this corresponds to
the \textit{past horizon }because along the outgoing trajectory (\ref{10}) the
relation between the Painleve-Gullstrand and Schwarzschild times following
from (\ref{6}) is $t_{p}=2t+const.$ Consequently, minus infinity in the
Painleve-Gullstrand time corresponds to minus infinity in the Schwarzschild
time. Also the action along the trajectory (\ref{10}) is regular and real, as
can be seen substituting expression (\ref{10}) for $t_{p}$ and expression
(\ref{9}) for $W_{p}(r)$ into the formula $S=-E_{p}t_{p}+W_{p}(r)$. This is
due to the fact that the Painleve-Gullstrand time $t_{p}$ is regular and real
only around the future horizon but in the vicinity of the past horizon this
variable is as \textquotedblleft bad" as the Schwarzschild time---no
difference. On the past horizon, $t_{p}$ goes to minus infinity and acquires
an imaginary additive term after crossing it, a term which cancels the
corresponding imaginary part of the function $W_{p}(r)$.

As for the future horizon we can gain nothing from the regularity and reality
of the Painleve-Gullstrand time in that region because no outgoing
trajectories exist which would be able to come into contact with the future
horizon. In other words \textit{to treat the expression }$S=-E_{p}t_{p}%
+W_{p}(r)$\textit{\ with real and regular }$t_{p}$\textit{\ and at the same
time with a pole singularity in }$dW_{p}/dr$\textit{\ as an action
corresponding to a particle trajectory is nonsense. Such a trajectory does not
exist, namely it does not satisfy the equation of motion }$-t_{p}%
+dW_{p}/dE_{p}=const.$ The main advantage of the Painleve-Gullstrand time
around points of the future horizon is irrelevant to the question of
tunneling.\footnote{An \textit{ingoing} particle can freely cross the future
horizon from outside to inside following the classical path. Because the
action $S(r)$ for such a path is regular and real and the Painleve-Gullstrand
variable $t_{p}$ is regular and real around the future horizon, the momentum
$dW_{p}/dr$ should not contain any pole singularity at points $r=r_{g}$. That
this is so can be seen from the formula (\ref{8}) since for ingoing particles
the sign of the square root inside the bracket of this formula should be taken
to be a minus sign (in this case in the region $r>r_{g}$ momentum
\newline$p_{p}(r)<0$ which correspond to the particles\textbf{ }coming from
infinity).}

In turn all of this discussion shows that for the evaluation the probability
$w\sim\exp\left(  -2|\operatorname{Im}\,\triangle S|/\hbar\right)  $ for
particle creation in a black hole space-time, one needs to bear in mind the
complete action $S$ (along that geodesic which is a candidate for the particle
creation trajectory) and not only its shortened part $W(r)$. Otherwise one can
obtain the mistaken result that particles can be created in empty flat space
just due to a different choice of coordinates. Indeed, take the metric
$-ds^{2}=-\left(  dx^{0}\right)  ^{2}+\left(  dx^{1}\right)  ^{2}$ and make
the following coordinate transformation:
\begin{equation}
x^{0}=2r_{g}\sqrt{\frac{r}{r_{g}}-1}\sinh\frac{ct}{2r_{g}}\text{ },\text{
\ }x^{1}=2r_{g}\sqrt{\frac{r}{r_{g}}-1}\cosh\frac{ct}{2r_{g}}\,. \label{11}%
\end{equation}
The resulting metric is $-ds^{2}=-\left(  r/r_{g}-1\right)  c^{2}%
dt^{2}+\left(  r/r_{g}-1\right)  ^{-1}dr^{2}$. Again, the action can be
written as $S=-Et+W(r)$ and from the Hamilton-Jacobi equation we have
$dW/dr=c^{-1}\left(  r-r_{g}\right)  ^{-1}\sqrt{E^{2}r_{g}^{2}-m^{2}c^{4}%
r_{g}\left(  r-r_{g}\right)  }$. If one insists that the action $S(r)$ along
some geodesic has an imaginary part produced by $W$ (due to the pole
singularity in the momentum $dW/dr$ at the horizon $r=r_{g})$, this would
result in an imaginary part of the action also in the Minkowski coordinates
$x^{0},x^{1}$ because the action along a geodesic is relativistically
invariant. To be saved from this error it is necessary to also take into
account the contribution from the term $Et$ in the action, a term which as
calculated along the particle trajectory also contains an imaginary part and
this part compensates for the imaginary additive term in the function $W(r)$
exactly so as to guarantee the reality of the action $S$. This is precisely
the same as what also happens in the vicinity of the horizon in the black hole space-time.

It is easy to recognize the basic role of the equivalence principle behind all
of the preceding discussion. We can formulate this in the following way.
\textit{The absence of any imaginary part in the action along an
infinitesimally small geodesic segment in any regular space-time neighborhood
is a general consequence of the equivalence principle. Any horizon point in
fact is a regular point of space-time in the vicinity of which the flat metric
can be introduced. In this locally flat neighborhood the equations of
geodesics and the Hamilton-Jacobi equation for the action are the same as in
flat Minkowski space-time because they depend only on metric and Christoffel
symbols and do not contain the second or higher order derivatives of the
metric tensor. Consequently in such a neighborhood the change of the
particle's action cannot acquire any imaginary part (if it did, particles
would be created in empty Minkowski space). Since the action }$S=-mc\int ds$
\textit{along a particle trajectory is a relativistically invariant quantity
it cannot have any imaginary part in any other coordinates (whether singular
or not) covering this neighborhood.}

To see the mathematical realization of this fundamental property of the
gravitational field in the case of a black hole consider any metric of the
form%
\begin{equation}
-ds^{2}=g_{00}(x^{1})(dx^{0})^{2}+g_{11}(x^{1})(dx^{1})^{2}+2g_{01}%
(x^{1})dx^{0}dx^{1}\,. \label{11-1}%
\end{equation}
The particle's action is%
\begin{equation}
S=-c^{-1}Ex^{0}+W(x^{1}) \label{11-2}%
\end{equation}
and from the Hamilton-Jacobi equation it follows that:%
\begin{equation}
\frac{dW}{dx^{1}}=-\frac{Eg_{01}+\varepsilon\sqrt{\left(  E^{2}+m^{2}%
c^{4}g_{00}\right)  \left(  -g\right)  }}{cg_{00}}\,, \label{11-3}%
\end{equation}
where $g=g_{00}g_{11}-g_{01}^{2}$ and $\varepsilon=\pm1$ ($\varepsilon$
distinguishes outgoing and ingoing trajectories). The equation of motion
$\partial S/\partial E=const.$ takes the form:%
\begin{equation}
x^{0}=-\int\left\{  \frac{g_{01}}{g_{00}}+\frac{\varepsilon E}{g_{00}%
\sqrt{\left(  E^{2}+m^{2}c^{4}g_{00}\right)  \left(  -g\right)  ^{-1}}%
}\right\}  dx^{1}+const\,. \label{11-4}%
\end{equation}
If we are interested in calculating the action along the particle trajectory
we have to substitute this expression for $x^{0}$ into the formula
(\ref{11-2}) which gives:%
\begin{equation}
S=-m^{2}c^{3}\varepsilon\int\frac{dx^{1}}{\sqrt{\left(  E^{2}+m^{2}c^{4}%
g_{00}\right)  \left(  -g\right)  ^{-1}}}\,. \label{11-5}%
\end{equation}
The expression (\ref{11-1}) covers in a unique way all three of the most
popular choices of coordinates for the \textquotedblleft time"-independent
form of the black hole metric, namely Schwarzschild, Eddington-Finkelstein and
Painleve-Gullstrand coordinates. For each of these cases the metric
determinant has the value $g=-1$ and the horizon has the equation $g_{00}=0.$
Then from (\ref{11-5}) we see that there are no poles on the horizon in the
integrand of the action. Around any horizon point the action is regular and
real and this fact does not depend on the type of the trajectory (outgoing or
ingoing) and on the type of the horizon (past or future). We repeat again that
this is the straightforward consequence of the equivalence principle and the
covariance of general relativity: any infinitesimal neighborhood of any
horizon point is homeomorphic to Minkowski space and the choice of coordinates
for the calculation of any relativistic invariant in such a neighborhood is of
no importance. In any coordinates an invariant like the imaginary part of the
action along a geodesic will be the same as in empty Minkowski space-time,
namely zero since in the empty Minkowski space-time there are no physical
obstacles which would be able to force a particle to jump into the complex
space-time domain.\footnote{This is true also in the Rindler coordinates of an
accelerated observer \textit{moving in the Minkowski space-time.} A reader
should not mix up this case with the quite different artificial models where
together with the coordinate transformation (\ref{11}) one introduces at the
horizon points some hidden physical obstacle or boundary condition for the
quantum field. This can happen if one uses a quantization procedure which in
fact is invalid in the Minkowski space-time but acceptable in the space-time
with the aforementioned boundary condition. Of course, in such artificial
constructions the \textquotedblleft horizon" points have no neighborhoods
homeomorphic to flat space-time and arguments based on the equivalence
principle and covariance cannot be applied. Such models correspond to the
so-called Hawking effect and the Unruh effect. However, these models have
nothing to do with black holes and accelerated systems which have horizons
free of any obstacles. For more details see our section \textquotedblleft
Discussion".}

It is worth adding the following remark in order to avoid misunderstanding,
the source of which is the confusion of notions between the case of a
stationary potential's non-relativistic quantum mechanics and the properties
of a black hole space-time. \textit{The Schwarzschild and Painleve-Gullstrand
\textquotedblleft time" variables have spacelike coordinate lines inside the
horizon and the conserved quantities }$E$\textit{\ and }$E_{p}$%
\textit{\ globally are not energies.} Therefore it would be a mistake to apply
to a black hole the tunneling approach coming from the standard
Schr\"{o}dinger theory for a stationary potential, since it makes no sense to
use the Schr\"{o}dinger equation with spacelike variables $t$ or $t_{p}$ in
place of the time inside the horizon. There do not exist coordinates with
which one can cover both sides of the horizon of a black hole in such a way
that one of these coordinates is globally timelike and the metric does not
depend on it. \textit{The black hole space-time is not only not static, it is
also not stationary independent of which system of reference we use, including
the Painleve-Gullstrand coordinates.} Then to be able to speak about quantum
mechanics and tunneling one needs to introduce first the coordinate which
would be everywhere regular and timelike both outside and inside of the
horizon. However, the metric in this case will be unavoidably time-dependent.
Therefore we are forced to the necessity of applying the quasi-classical
tunneling approach to an essentially time-dependent external gravitational
field which is quite different from the naive extrapolation of the
non-relativistic theory for the stationary potential proposed in
\cite{V}--\cite{CHNVZ}. The calculation of the action $S=-mc\int ds$ along a
quasi-classical complex trajectory in a black hole space-time must take into
account in that or another way the contribution from all variables
involved.\footnote{It is reasonable to note that another specific criticism
(quite different from that presented here) of the tunneling method as
described in \cite{PW} has been stated by B. Chowdhury \cite{Ch}. In spite of
some attempts that have appeared in the literature to overcome the difficulty
indicated in \cite{Ch}, the problem remains unsolved and points to still
another type of inconsistency of the tunneling approach proposed in the
article \cite{PW}.}

\section{On tunneling in the space-time of a physical black hole}

The non-existence of any tunneling through an \textit{infinitesimal
}neighborhood of a black hole horizon does not yet give grounds to assert that
no tunneling exists at all. The question still remains whether particles can
be created by a black hole gravitational field in a \textit{finite space-time
region}, a problem which is essentially different with respect to the previous
discussion. The best way to study this problem is to use the Kruskal
coordinates from the outset. This was done in \cite{B3} and the result was
that \textit{in the space-time of a physical black hole formed by collapse, no
quasi-classical tunneling trajectories between the inside and outside of
horizon can be found independent of the size of the assumed tunneling domain}.

Let's reproduce some of the principal results of reference \cite{B3}. First we
recall how the quasi-classical particle creation process can be described in a
non-gravitational setting. When a particle moves in an external field in the
flat space-time of the metric $-ds^{2}=-c^{2}dt^{2}+dx^{2}$ we have the
following integral of the motion:
\begin{equation}
\left(  \frac{cdt}{ds}\right)  ^{2}=1+\left(  \frac{dx}{ds}\right)  ^{2}\,,
\label{12}%
\end{equation}
where $t=t(s,C_{1},C_{2})$, $x=x(s,C_{1},C_{2})$ are parametric equations of
the trajectories in which any fixed values of the constants $C_{1},C_{2}$ pick
out a particular trajectory. The barrier between two Dirac seas (in one of
which $cdt/ds>0$ and in the other $cdt/ds<0$) shows itself as the cut along
the segment $(-i,i)$ of the imaginary axis of the complex plane of the
variable $dx/ds.$

If we introduce the parameter $R$ by the relations
\begin{equation}
\frac{cdt}{ds}=\cosh R\text{ },\text{ \ }\frac{dx}{ds}=\sinh R\text{
}\,,\label{13}%
\end{equation}
then any quasi-classical path (if it exists, of course, that is if all other
equations of motion, and not only the integral (\ref{12}), permit such
trajectories) making a contribution to the pair creation probability
corresponds to the continuous contour $ABCD$ in the complex plane of the
variable $R$, shown in Fig.1. The quantity $n$ in that figure can be any
integer including zero and the main contribution to the probability of the
process comes from the value $n=0$. In the sequel we consider $n$ to be
non-negative and this does not restrict the generality of our
analysis.\footnote{By the way we would like to note that In the article
\cite{B3} the contour in the complex plane of the variable $dx/ds$ shown there
by its Fig.~2 in fact does not corresponds to the path $ABCD$ shown in Fig.~3
of article \cite{B3} and by Fig.~1 of the present article. This was a misprint
but it changes nothing in the results of the article \cite{B3}, because its
Fig.2 has not been used for any calculation made in paper \cite{B3}. In that
paper the correct principal ($n=0$) particle creative contour in the complex
plain of $dx/ds$ has to go from one Riemann sheet of the function $cdt/ds$ to
the other around the branch points $dx/ds=i$ in such away that $dx/ds$ is
positive both before and after sub-barrier transition.} The path shown in
Fig.1 contains the sub-barrier jump represented by the vertical purely
imaginary part $BC$ of the contour, after which the time component $cdt/ds$ of
the relativistic velocity changes sign. For example, for the case of a
constant electric field the trajectories which realize the contour $ABCD$
indeed exist and the imaginary part of the action (for each $n$) along the
under-barrier transition $BC$ coincides with that one which follows from the
exact Schwinger theory of electron-positron pair creation.%
\begin{figure}
[ptb]
\begin{center}
\includegraphics[
trim=0.000000in 0.000000in -0.022485in 0.002488in,
height=2.1006in,
width=2.5944in
]%
{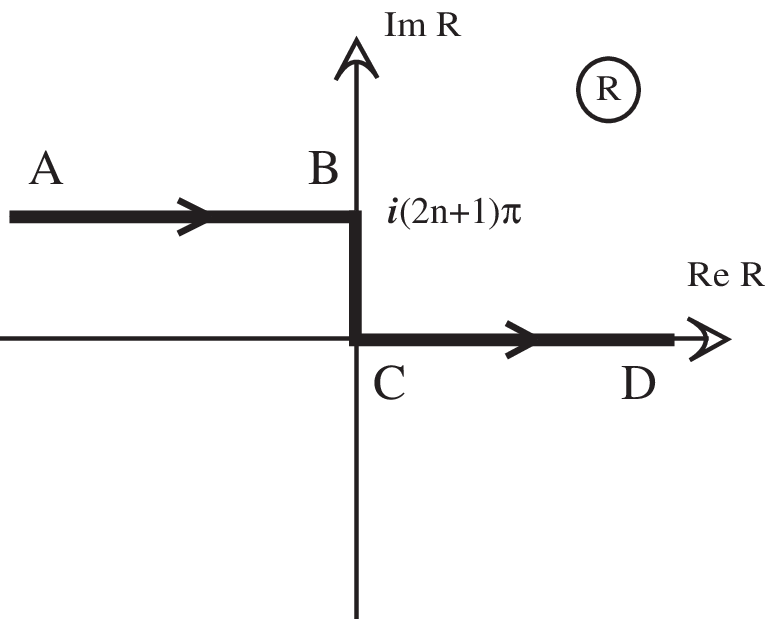}%
\caption{The particle creation contour in the complex plane of $R$.}%
\label{fig1}%
\end{center}
\end{figure}

The generalization to a space-time with the metric $-ds^{2}=g_{ik}dx^{i}%
dx^{k}$ is straightforward and can be done in a covariant way. For a particle
moving in such a space-time we have the same type of integral of the motion
$g_{ik}(dx^{i}/ds)(dx^{k}/ds)=-1$ which in the two-dimensional case (which we
consider only for simplicity, the same approach has an obvious extension for
any dimension) can be written in the following form:%
\begin{equation}
\left(  l_{i}\frac{dx^{i}}{ds}\right)  ^{2}=1+\left(  m_{i}\frac{dx^{i}}%
{ds}\right)  ^{2}, \label{14}%
\end{equation}
where ($i,k=0,1$) and where we use an orthonormal frame consisting of two
vectors $l_{i},m_{i}$ in terms of which the metric tensor can be represented
as $g_{ik}=-l_{i}l_{k}+m_{i}m_{k}$. We have to construct $l_{i}$\textit{\ as a
globally continuous future-directed timelike vector}. We can similarly
introduce the parameter $R$ in this space-time:%
\begin{equation}
-l_{i}\frac{dx^{i}}{ds}=\cosh R\text{ },\text{ \ }m_{i}\frac{dx^{i}}{ds}=\sinh
R\,, \label{15}%
\end{equation}
and the barrier between two Dirac seas (in one of which $l_{i}dx^{i}/ds>0$ and
in the other $l_{i}dx^{i}/ds<0$) shows itself as the cut along the segment
$(-i,i)$ of the imaginary axis of the complex plane of the variable
$m_{i}dx^{i}/ds$. In order for particle creation to take place, there should
exist a complexified geodesic trajectory corresponding to the same contour as
in Fig.1. Globally such a trajectory can be described by the equations
$x^{0}=x^{0}(s,C_{1},C_{2}),$ $x^{1}=x^{1}(s,C_{1},C_{2})$ with some fixed
values of the arbitrary constants $C_{1,}C_{2}$. The four crucial properties
(which represent the direct generalization of what is going in the flat
space-time) of such a particle creation geodesic should be: \textit{(i) the
relevant three parts of such a trajectory, the one corresponding to the
under-barrier path }$BC$\textit{\ and to the two regular segments of contours
}$AB$\textit{\ and }$CD$ \textit{(or to the entire half-lines }$AB$%
\textit{\ and }$CD$\textit{\ if they are free of singularities) adjacent to
}$BC,$ \textit{represent an analytical continuation of each other with respect
to the proper time }$s$, \textit{(ii) both the classical initial part of }%
$AB$\textit{\ (adjoining the point }$B$) \textit{and final part of }%
$CD$\textit{\ (adjoining the point }$C$\textit{) of this trajectory belong to
the same space-time,} \textit{(iii) the projection }$l_{i}dx^{i}%
/ds$\textit{\ of the relativistic velocity onto the time-like leg of the
orthogonal frame changes sign after the sub-barrier transition }$BC$
\textit{(the sign of the quantity }$l_{i}dx^{i}/ds$\textit{\ is invariant with
respect to both coordinate transformations and proper orthochronous Lorentz
rotations of the frame, hence this sign represents a physical quantity with
which one can distinguish between particle and antiparticle), (iiii) the
space-time separation between the creation points }$B$\textit{\ and }%
$C$\textit{\ must be space-like, i.e., each of these points should be located
outside of the light cone of the other (only under this condition the
appearance of particles at points }$B$\textit{\ and }$C$\textit{\ can be
interpreted as creation).}

Due to tunneling, the change $\mathbf{\triangle}s$ of the proper time $s$
along the trajectory gains an extra imaginary additive term $\operatorname{Im}%
\mathbf{\triangle}s$. Because the particle's action in the gravitational field
is $S=-mc\int ds$, this imaginary part defines the order of magnitude of the
main contribution to the probability $w$ of the process by means of the
formula
\begin{equation}
w\sim\exp\left(  -2mc|\operatorname{Im}\,\triangle s|/\hbar\right)  \,.
\label{16}%
\end{equation}
The imaginary part of the difference of $R$ between the final and initial
classical sectors of the trajectory is $\triangle R=i\left(  2n+1\right)  \pi$
and one can find $\operatorname{Im}\triangle s$ if the relation between the
parameters $R$ and $s$ is known. In this way one can get the main exponential
factor in the probability for the process in accordance with the formula
(\ref{16}).

Let us apply this approach to the black hole space-time covered by the Kruskal
coordinates $\eta$ and $\zeta$ which are related to the Schwarzschild
coordinates of the metric (\ref{1}) by the transformation:
\begin{equation}
r^{\ast}+ct=2r_{g}\ln(\zeta+\eta)\,,\text{ }r^{\ast}-ct=2r_{g}\ln(\zeta
-\eta)\,,\text{ }r^{\ast}=r+r_{g}\ln(\frac{r}{r_{g}}-1)\,.\text{ } \label{17}%
\end{equation}
The Kruskal form of the metric is:%
\begin{equation}
-ds^{2}=\frac{4r_{g}^{3}}{r}e^{-\frac{r}{r_{g}}}(-d\eta^{2}+d\zeta
^{2})\,,\quad\label{18}%
\end{equation}
where $r\left(  \eta,\zeta\right)  $ follows from the relation:%
\begin{equation}
\eta^{2}-\zeta^{2}=\left(  1-\frac{r}{r_{g}}\right)  e^{\frac{r}{r_{g}}}\,.
\label{18-1}%
\end{equation}
Without loss of generality we can choose any orthogonal frame with globally
continuous vectors $l_{i}$ and $m_{i}$ (apart from points at the singularity
$r=0$). The simplest choice is%
\begin{equation}
l_{i}=(-\sqrt{\omega},0);\quad\quad m_{i}=(0,\sqrt{\omega})\,, \label{19}%
\end{equation}
where%
\begin{equation}
\omega=\frac{4r_{g}^{3}}{r}e^{-r/r_{g}}\,. \label{20}%
\end{equation}

The real space-time of interest with metric (\ref{18}) covered by the real
coordinates $\eta$ and $\zeta$ represents one sheet on which $r>0$ everywhere.
Therefore we can define the \textit{single-valued positive functions }%
$\sqrt{r}$\textit{\ and }$\sqrt{\omega}$ on this sheet. Such a choice conforms
to the condition that the timelike vector $l_{i}$ in (\ref{19}) should be
everywhere directed into the future. Now the integral (\ref{14}) expressed
through (\ref{15}) takes the form:%
\begin{equation}
\sqrt{\omega}\,\frac{d\eta}{ds}=\cosh R,\quad\quad\sqrt{\omega}\,\frac{d\zeta
}{ds}=\sinh R \label{21}%
\end{equation}
and it is easy to show that the geodesic equations for the metric (\ref{18})
allow one more integral:%
\begin{equation}
\omega\zeta\frac{d\eta}{ds}-\omega\eta\frac{d\zeta}{ds}=\frac{2r_{g}E}{mc^{2}%
}\,. \label{22}%
\end{equation}
where $E$ is an arbitrary constant (the conserved Schwarzschild energy). Both
results (\ref{21}) and (\ref{22}) follow from the known first integrals of the
geodesic equations in the Schwarzschild coordinates (that is $\left(
dr/ds\right)  ^{2}=\left(  E/mc^{2}\right)  ^{2}-1+r_{g}/r$ and
$cdt/ds=\left(  E/mc^{2}\right)  \left(  1-r_{g}/r\right)  ^{-1}$) after we
transform them to the Kruskal variables $\eta,\zeta$ in accordance with
(\ref{17}).

The general solution of Eqs.~(\ref{21}) and (\ref{22}) can be written in
parametric form with the aid of the auxiliary parameter $\lambda$:
\begin{equation}
\zeta=\left(  \frac{\sqrt{1+\alpha^{2}}}{\alpha}\sinh\lambda\cosh
R-\cosh\lambda\sinh R\right)  \exp\left(  \frac{\sinh^{2}\lambda}{2\alpha^{2}%
}\right)  \,, \label{24}%
\end{equation}%
\begin{equation}
\eta=\left(  \frac{\sqrt{1+\alpha^{2}}}{\alpha}\sinh\lambda\sinh
R-\cosh\lambda\cosh R\right)  \exp\left(  \frac{\sinh^{2}\lambda}{2\alpha^{2}%
}\right)  \,, \label{25}%
\end{equation}
where the constant $\alpha$ is defined by the relation:%
\begin{equation}
\alpha^{2}=\frac{E^{2}-m^{2}c^{4}}{m^{2}c^{4}} \label{26}%
\end{equation}
and the variable $R$, the proper time $s$, and the Schwarzschild radial
coordinate $r$ can be expressed as functions of the same parameter $\lambda$:%
\begin{equation}
R=\frac{\sqrt{1+\alpha^{2}}}{4\alpha^{3}}\left[  \sinh(2\lambda)+2\left(
2\alpha^{2}-1\right)  \lambda\right]  -Q\,,\text{ } \label{27}%
\end{equation}%
\begin{equation}
s=\frac{r_{g}}{2\alpha^{3}}[\sinh(2\lambda)-2\lambda]\,, \label{28}%
\end{equation}%
\begin{equation}
\sqrt{\frac{r}{r_{g}}}=\frac{\sinh\lambda}{\alpha}\,. \label{29}%
\end{equation}
The quantity $Q$ in (\ref{27}) is another arbitrary constant, which represents
the new integral of motion appearing in the course of the integration of
Eqs.~(\ref{21}) and (\ref{22}). (On the left hand side of Eq.~(\ref{28}) an
arbitrary additive constant should appear but it can be taken to be zero
without loss of generality.) \footnote{The special case $\alpha=0$ also can be
treated in the framework of the same expressions (\ref{24})--(\ref{29}) by the
parameter redefinition $\lambda=\alpha\acute{\lambda}.$ After that all
formulas will have a well defined limit at $\alpha\rightarrow0.$}

Most interesting for us is the case of real constant $\alpha$ corresponding to
those trajectories having $E>mc^{2}$ which can escape to spatial infinity.
Then in the sequel we will study exactly this case (in the context of particle
creation the case $E<mc^{2}$ is irrelevant not merely because such particles
cannot escape to infinity but because creation of such trapped particles is
impossible to begin with, see \cite{B3}).

Consider a complexified geodesic which corresponds to some fixed values of
constants $\alpha$ and $Q$ and which we consider as a candidate for a particle
creation trajectory. If the variable $R$ evolves along the sub-barrier segment
$BC$ we can find out from (\ref{27}) the corresponding evolution of the
parameter $\lambda=\operatorname{Re}\lambda+i\operatorname{Im}\lambda$ and, in
particular, the values of $\lambda$ at points $B$ and $C,$ where particles are
coming into being. Because the quantity $Q$ is a constant of motion and
$\operatorname{Re}R$ is zero at points $B$ and $C$ we have to equate the
values of the expression $\operatorname{Re}[\sinh(2\lambda)+2\left(
2\alpha^{2}-1\right)  \lambda]$ at points $B$ and $C$. At first, taking into
account that at these points $\sqrt{r}$ is real and recalling the elementary
formulas:%
\begin{align}
\sinh\left(  f+ig\right)   &  =\sinh f\cos g+i\cosh f\sin g\,,\label{29-1}\\
\cosh\left(  f+ig\right)   &  =\cosh f\cos g+i\sinh f\sin g\,,\nonumber
\end{align}
we deduce from (\ref{29}) that
\begin{equation}
\lbrack\sin\left(  \operatorname{Im}\lambda\right)  ]_{B}=[\sin\left(
\operatorname{Im}\lambda\right)  ]_{C}=0\,. \label{29-2}%
\end{equation}
Then we obtain:%
\begin{equation}
\left[  \sinh(2\operatorname{Re}\lambda)+2\left(  2\alpha^{2}-1\right)
\operatorname{Re}\lambda\right]  _{B}=\left[  \sinh(2\operatorname{Re}%
\lambda)+2\left(  2\alpha^{2}-1\right)  \operatorname{Re}\lambda\right]
_{C}\,. \label{30}%
\end{equation}
If $\alpha^{2}>0$ the equation (\ref{30}) has only one solution:%
\begin{equation}
\left(  \operatorname{Re}\lambda\right)  _{B}=\left(  \operatorname{Re}%
\lambda\right)  _{C}\,. \label{31}%
\end{equation}
With this result, taking into account the condition that $\sqrt{r}$ at points
$B$ and $C$ must be not only real but also positive, we see immediately from
(\ref{29}) and (\ref{29-2}) that either%
\begin{equation}
\left(  \operatorname{Re}\lambda\right)  _{B}=\left(  \operatorname{Re}%
\lambda\right)  _{C}>0,\text{ }[\cos\left(  \operatorname{Im}\lambda\right)
]_{B}=[\cos\left(  \operatorname{Im}\lambda\right)  ]_{C}=1\,, \label{31-1}%
\end{equation}
or%
\begin{equation}
\left(  \operatorname{Re}\lambda\right)  _{B}=\left(  \operatorname{Re}%
\lambda\right)  _{C}<0,\text{ }[\cos\left(  \operatorname{Im}\lambda\right)
]_{B}=[\cos\left(  \operatorname{Im}\lambda\right)  ]_{C}=-1\,. \label{31-2}%
\end{equation}
However, in both cases $\sinh\lambda$ and $\cosh\lambda$ are real at points
$B$ and $C$ and, keeping in mind the condition (\ref{31}), we obtain:%
\begin{equation}
\left(  \sinh\lambda\right)  _{B}=\left(  \sinh\lambda\right)  _{C}\text{
},\text{ }\left(  \cosh\lambda\right)  _{B}=\left(  \cosh\lambda\right)
_{C}\,. \label{32}%
\end{equation}

Now we can get the values of the Kruskal coordinates at points $B$ and $C$.
Because $\left(  \sinh R\right)  _{B}=0$ $,$ $\left(  \cosh R\right)  _{B}=-1$
and $\left(  \sinh R\right)  _{C}=0$ $,$ $\left(  \cosh R\right)  _{C}=1$ from
(\ref{24}) and (\ref{25}) follows the main result:%
\begin{equation}
\left(  \zeta,\eta\right)  _{B}=\left(  -\zeta,-\eta\right)  _{C}\text{ }.
\label{33}%
\end{equation}
This shows that a particle in the external field described by the metric
(\ref{18}) could perform the tunneling from some point ($\eta,\zeta$) only to
the point ($-\eta,-\zeta$), that is \textit{the creation points should be
reflections of each other across the origin of the Kruskal coordinate system}.
\textit{However, in case of a black hole formed by the collapse, in the domain
of interest, that is in the empty space outside the collapsing matter, no such
pair of points exists.} This is the result which confirms our claim made in
the first paragraph of this section.

\section{On tunneling in the space-time of an eternal black hole}

Until now eternal black holes have no physical applications so the problem of
tunneling in their gravitational field can only be of academic interest.
Nevertheless, for the formal completeness of our study we also present the
analysis of this case.

If we have an eternal black hole then the conjugate points $(\zeta,\eta)$ and
$\left(  -\zeta,-\eta\right)  $ are at our disposal and it might seem that the
appearance of a particle creation effect can be present. However, it is
important to realize that Eq.~(\ref{33}) represents only the necessary
condition which by no means guarantees that a continuous sub-barrier geodesic
connecting such conjugate points indeed exists. In fact it turns out that
\textit{even} \textit{in the space-time of an eternal black hole no
sub-barrier path which would be able to connect points }$(\zeta,\eta
)$\textit{\ and }$\left(  -\zeta,-\eta\right)  $ \textit{and which can be
interpreted as a particle creation trajectory exists.}

To find the complete set of existence conditions for the particle creation
trajectories let's return to the integral of motion (\ref{27}) and examine how
the parameter $\lambda$ changes when $R$ goes along the segment $BC$. Tracing
out separately the real and imaginary parts of the relation (\ref{27}) we can
obtain equations for those trajectories in the complex plane of $\lambda$
which correspond to a path along the imaginary axes of the complex plane of
the parameter $R$. With the notation%
\begin{equation}
R=iv\,,\text{ }2\lambda=x+iy \,, \label{34}%
\end{equation}
these equations take the form:%
\begin{equation}
\sinh x\cos y+\left(  2\alpha^{2}-1\right)  x=\frac{4\alpha^{3}\left(
\operatorname{Re}Q\right)  }{\sqrt{1+\alpha^{2}}}\,, \label{35}%
\end{equation}%
\begin{equation}
\cosh x\sin y+\left(  2\alpha^{2}-1\right)  y=\frac{4\alpha^{3}\left(
v+\operatorname{Im}Q\right)  }{\sqrt{1+\alpha^{2}}}\,. \label{36}%
\end{equation}

If we differentiate the last two relations with respect to the variable $v$ we
obtain the two-dimensional dynamical system $dx/dv=f_{1}\left(  x,y,\alpha
\right)  ,$ $dy/dv=f_{2}\left(  x,y,\alpha\right)  $ depending only on one
external parameter $\alpha.$ In the range of \textit{positive }$\alpha$
(negative $\alpha$ introduces nothing new, this case can be obtained from the
case of positive $\alpha$ by the reflection $x\rightarrow-x,$ $y\rightarrow
-y$) this system undergoes a mild change after crossing the value
$\alpha=1/\sqrt{2}$ but bifurcates essentially if one crosses the point
$\alpha=1$. Then to study the behavior of the trajectories the range of values
of $\alpha$ should be divided into the three qualitatively different regions:
$0<\alpha<$ $1/\sqrt{2},$ $1/\sqrt{2}<\alpha<$ $1$ and $\alpha>1$. The phase
portraits of this dynamical system for each of these domains of the values of
$\alpha$ we demonstrate in Fig.2. We show the integral curves only in the
principal interval $\left[  0,2\pi\right]  $ of the variable $y$. Due to the
periodicity of the system in $y$ these diagrams can be used as basic building
blocks for constructing the complete phase portrait for any interval of $y$.%
\begin{figure}
[ptb]
\begin{center}
\includegraphics[
height=4.74in,
width=1.9873in
]%
{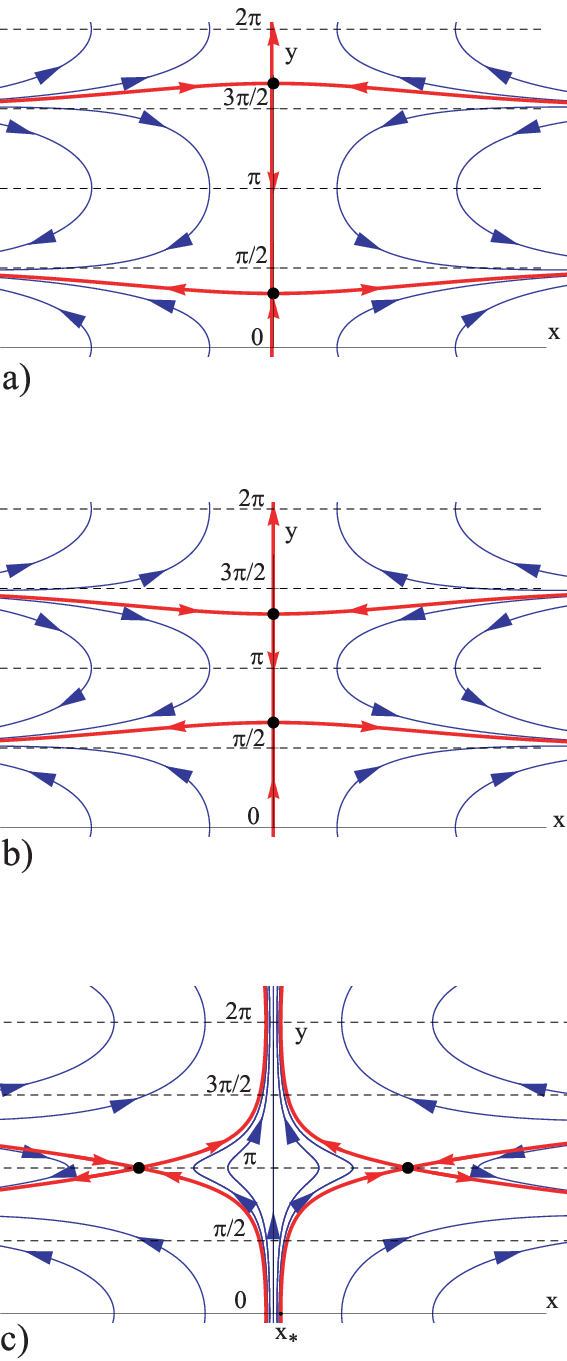}%
\caption{Trajectories $x=x(v,\alpha,Q),$ $y=y(v,\alpha,Q)$ of the dynamical
system (44)-(45) for real and positive constants $\alpha$ and $Q.$ Arrows
correspond to the growth of the parameter $v.$ a) the case $0<\alpha
<1/\sqrt{2}.$ In the main interval $[0,2\pi]$ there are two saddle points
$y=y_{1}$ $(0<y_{1}<\pi/2)$ and $y=y_{2}$ $(3\pi/2<y_{2}<2\pi)$ on the axis
$x=0$ where $y_{1}$ and $y_{2}$ are two solution of the equation $\cos
y=1-2\alpha^{2},$ and all saddle's separatrices are described by the equation
(44) at $Q=0$. b) the case $1/\sqrt{2}<\alpha<1.$ Qualitatively everything is
the same as on the previous diagram. The difference is that the $y$%
-coordinates of the saddles are drawn together, now they are closer to the
center $(\pi/2<y_{1}<\pi,$ $\pi<y_{2}<3\pi/2)$. c) the case $\alpha>1.$ The
system bifurcates so that the channel for the sub-barrier paths appeared. Each
outgoing trajectory which starts at $(x=x_{C,}$ $y=0)$ where $0<x_{C}<x_{\ast
}$ can pass through this channel. The initial value $x_{C}>0$ of the
$x-$coordinate is defined by an arbitrary constant $Q>0$ and follows from the
equation $4\alpha^{3}Q=\sqrt{1+\alpha^{2}}\left[  \sinh x_{C}+\left(
2\alpha^{2}-1\right)  x_{C}\right]  $. The location of point $x_{\ast}$ is
defined by the relation $\sinh x_{\ast}+\left(  2\alpha^{2}-1\right)  x_{\ast
}=2\left(  2\alpha^{2}-1\right)  \ln\left(  \alpha+\sqrt{\alpha^{2}-1}\right)
-2\alpha\sqrt{\alpha^{2}-1}.$ Now in the right half-plane of the phase diagram
we have saddle point $(x=x_{s},$ $y=\pi),$ where $x_{s}>0$ has to be
calculated from the relation $\sinh x_{s}=2\alpha\sqrt{\alpha^{2}-1}.$ The
equations for corresponding separatrices are $\sinh x\cos y+\left(
2\alpha^{2}-1\right)  x=2\left(  2\alpha^{2}-1\right)  \ln\left(  \alpha
+\sqrt{\alpha^{2}-1}\right)  -2\alpha\sqrt{\alpha^{2}-1}.$}%
\label{fig2}%
\end{center}
\end{figure}

It follows from Eqs.~(\ref{28})--(\ref{29}) that $dr/ds=\alpha\coth\lambda$
and it is easy to see that the sign of $\operatorname{Re}\lambda$ defines the
outgoing-ingoing type of a trajectory. Because we choose the constant $\alpha$
to be positive and on the classical branches of any trajectory $\sqrt{r}$ must
be real and positive, from (\ref{29}) it results that on the classical parts
of the trajectories we have either $\lambda=\operatorname{Re}\lambda+2iN\pi$,
$\operatorname{Re}\lambda>0$ or $\lambda=\operatorname{Re}\lambda+i\left(
2N+1\right)  \pi$, $\operatorname{Re}\lambda<0$ (in both cases $N$ takes the
integer values including zero). In the former case $\coth\lambda$ is real and
positive, then $dr/ds$ is real and positive which corresponds to the outgoing
geodesics. In the second case $\coth\lambda$ is real and negative, then
$dr/ds$ is real and negative which is in keeping with ingoing geodesics. Since
each of the sub-barrier paths should have the end points $\lambda=\lambda_{B}$
and $\lambda=\lambda_{C}$ in which, in accordance with (\ref{31}) satisfies
$\left(  \operatorname{Re}\lambda\right)  _{B}=\left(  \operatorname{Re}%
\lambda\right)  _{C}$, the right halves of the phase portraits shown in Fig.2
describe the behavior of the under-barrier complex parts of the outgoing
trajectories and the left halves of the diagrams correspond to the
under-barrier complex parts of the ingoing trajectories. These phase diagrams
show that no trajectory can pass from the right domain $\operatorname{Re}%
\lambda>0$ to the left domain $\operatorname{Re}\lambda<0$ or vice versa. This
means that any complexified quasiclassical geodesic has the well-defined
conserved \textquotedblleft quantum number" $sign\left(  \operatorname{Re}%
\lambda\right)  $ and each keeps the property to be outgoing or ingoing during
the course of the whole evolution including the complex sub-barrier
transition.\footnote{The equivalent outgoing-ingoing classification can be
obtained if we permit the constant $\alpha$ to take both signs but we will
keep $\operatorname{Re}\lambda$ to only be positive or only be negative.}

From the relations (\ref{31-1}), (\ref{31-2}) and (\ref{34}) follows that in
general the interval in the variable $y$ between the end points $B$ and $C$ of
the tunnel should take one of the following values%

\begin{equation}
y_{B}-y_{C}=4\pi N\text{ },\text{ }N=\pm1,\pm2,... \label{36-1}%
\end{equation}
However, \textit{for values of} $\alpha$\textit{\ between }$0$\textit{\ and
}$1$\textit{\ there is not a single sub-barrier trajectory which would be able
to connect such end points. }In order to do such a job a trajectory, as can be
seen from (\ref{31}), has to go from a point with $(x_{C},$ $y_{C})$ to the
point with the same $x$-coordinate $x_{B}=x_{C}$ and with $y_{B}=4N\pi+y_{C}$,
i. e. in the $y$-directions such a trajectory has to overcome a distance
larger than $4\pi.$ But any such a path is blocked by the horizontal
separatrices coming from (or incoming to) the saddle points on the Fig.2a and Fig.2b.

When we pass to the values $\alpha>1$ the system bifurcates so that the
saddles go away from the axes $y$ and their separatrices transform in such a
way that the channel for a bunch of continuous trajectories connecting the
points satisfying condition (\ref{36-1}) opens.\footnote{Only some discrete
set of values of $\alpha>1$ is permissible. From (\ref{36}), (\ref{36-1}) and
the fact that the change of parameter $v$ between points $B$ and $C$ is equal
to $(2n+1)\pi$ follows that $(2n+1)\alpha^{3}=\sqrt{1+\alpha^{2}}\left(
2\alpha^{2}-1\right)  N.$ This relation is the cubic equation with respect to
$\alpha^{2}.$ Its unique real root satisfying the condition $\alpha^{2}>1$
exists if and only if $2n+1<2N<\sqrt{2}(2n+1).$ The last inequality give all
possible allowed pairs of the (positive) integers $n$ and $N.$ For $n=0$ there
are no solutions at all since no integer $2N$ exists in the range between $1$
and $\sqrt{2}.$ The next possibility is $n=1$ in which case only one
acceptable value $N=2$ arises. With $n$ increasing more and more available
values of $N$ are coming into play. Then all permissible discrete values of
$\alpha^{2}$ (that is the discrete values of the Schwarzschild energy) come as
solutions of the aforementioned cubic equation for all permissible pairs of
integers $n$ and $N.$} Nonetheless, in spite of the existence of such
tunneling geodesics we cannot impart to them the status of particle creation
trajectories because they do not satisfy the fourth condition from the list of
requirements we enumerated in the second section of our paper: in this case
the end points of the tunneling paths have a time-like separation instead of
the necessary space-like one. It is easy to show that in the case $\alpha>1$
the following restrictions on the locations of the points $B$ and $C$ are
unavoidable:
\begin{equation}
\left(  \eta_{B}\right)  ^{2}-\left(  \zeta_{B}\right)  ^{2}>0,\text{ }\left(
\eta_{C}\right)  ^{2}-\left(  \zeta_{C}\right)  ^{2}>0\,, \label{36-2}%
\end{equation}
that is both of these points are located inside the light cone of the origin
of the Kruskal coordinates. Since $\eta_{B}=-\eta_{C}$ and $\zeta_{B}%
=-\zeta_{C}$ one of the ends of the tunnel is located inside the future cone
(that is inside the future horizon of the "black hole" region) and another one
inside the past cone (that is inside the past horizon of the "white hole"
region). This means that the separation between the points $B$ and $C$ is
time-like. \footnote{The proof is going as follows. Without loss of generality
we can consider a trajectory which starts from the point $C$ with coordinates
$(0<x_{C}<x_{\ast},$ $y_{C}=0).$ From (\ref{24})--(\ref{25}) it follows that
$sign\left[  \left(  \eta_{C}\right)  ^{2}-\left(  \zeta_{C}\right)
^{2}\right]  =sign\left(  \alpha^{2}-\sinh^{2}\lambda_{C}\right)  .$ At the
point $C$ the value of $\lambda_{C}=x_{C}/2$ should not exceed $x_{\ast}/2.$
Simple calculations, taking into account the defining equation for $x_{\ast}$
(see the caption to the Fig.2) show that if $\alpha>1$ this restriction leads
to the positiveness of the expression $\alpha^{2}-\sinh^{2}\lambda_{C}.$ Since
$\eta_{B}=-\eta_{C}$ and $\zeta_{B}=-\zeta_{C}$ we obtain the inequalities
(\ref{36-2}) for both end points of the tunnel.}

By the way it is worth stressing that the coordinate separation between the
ends of the tunnel cannot be arbitrarily small (which fact again demonstrates
the action of the equivalence principle). It follows from Eq.~(\ref{25}) that
the absolute value of the coordinate $\eta$ at points which correspond to the
points $B$ and $C$ of the contour $R$ is:%
\begin{equation}
\left\vert \eta\right\vert _{B,C}=\left\{  [\cosh(\operatorname{Re}%
\lambda)]\exp\left(  \frac{\sinh^{2}\left(  \operatorname{Re}\lambda\right)
}{2\alpha^{2}}\right)  \right\}  _{B,C}\,, \label{36-3}%
\end{equation}
and both of these values are greater then unity. Then for any pair of the end
points of the tunnel one of them is located in the upper part of the Kruskal
diagram between the horizontal line $\eta=1$ and upper branch of the
singularity curve $\eta^{2}-\zeta^{2}=1$ and the second in the lower part of
the diagram between the horizontal line $\eta=-1$ and the lower branch of the
singularity curve.

In article \cite{B3} the direct mathematical analysis of the existence of
tunneling geodesics between the points $B$ and $C$ for an eternal black hole
had not been performed. Instead a point of view was proposed that even if they
exist, the probability of such a tunneling should be considered to be zero.
Now we see that the problem is essentially simpler: such geodesics for the
case $0<\alpha<1$ are completely absent and for the case $\alpha>1$ they are
irrelevant to the effect of particle creation. Which kind of interpretation
(if any) can be found for the last case remains to be seen.

\section{Discussion}

In relation to the results of the preceding sections the natural question
arises how one can reconcile the absence of the effect of the quasiclassical
particle creation by a Schwarzschild black hole and the well known traditional
statements on the existence of this effect in the framework of the exact
second quantization approach. This problem has been investigated in \cite{B3}
and the answer is that also in the exact theory no flux of particles can arise
from a physical Schwarzschild black hole at asymptotically late times. Then
the absence of quasiclassical tunneling in fact is in agreement with the exact
theory. The point is that the traditional quantization procedure is applicable
if and only if the initial (that is before the onset of collapse, when a
collapsing star has infinite radius and zero density and the whole space-time
is flat) state of a quantum field satisfies the zero boundary condition on the
\textit{last ray} (the continuation of the future horizon into the infinite
past). This is the same type of the boundary condition which one must apply at
the points of horizon of an accelerated observer moving in the Minkowski
space-time in order to attribute a physical and mathematical sense to the
conventional quantization procedure based on the left and right Unruh modes
\cite{NFKMB1,NFKMB2,F1}. The conventional quantization in the black hole
space-time uses the modes which at the initial phase of collapse are the same
left and right modes but now with respect to the last ray. Such modes form an
incomplete set in the Minkowski space-time of the initial phase of the
collapse and they can be treated as complete only for a quite special
situation, namely when the quantized field vanishes at the points of the last
ray (or, equivalently, when the last ray is excised from the
space-time).\footnote{The problem here is that an arbitrary field cannot be
represented as a superposition of the left and right modes. The complete boost
mode \newline set (the set of eigenfunctions of the Lorentz boost operator),
apart from the left and right components, contains also the so-called
\textquotedblleft zero mode" which in principle has no decomposition in terms
of the left and right excitations. The crucial point is that this singular
\textquotedblleft zero mode" represents an essential part of the field since
it has nonzero measure in the space of the degrees of freedom of all physical
fields. All traditional calculations ignore the presence of the
\textquotedblleft zero mode" which is equivalent to the loss of an essential
set of degrees of freedom of the field, that is equivalent to the replacement
of the original system under examination by a quite different object.} This
means that in the framework of the standard approach the initial state of the
field never can be chosen as the Minkowski vacuum because such a boundary
condition contradicts the translational invariance of this vacuum. The initial
state with vanishing field along the last ray represents a condensate which is
packed by an infinite number of particles falling from spatial infinity mainly
along the last ray. Namely these particles which already existed at the onset
of collapse leak out into the external space at late times (in spite of the
infinite red shift because of the infinite energy density of these particles
in the vicinity of the last ray). Thus, there is not any process of late time
particle production and at the end of the collapse one observes only those
particles which have already been stored in the system from the beginning. The
same is true with regard to the thermal spectrum of the emergent flux: this
spectrum is a direct result of the Umezawa statistical noise \cite{Um} which
is present in the system from the beginning due to the peculiarity of the
initial conditions. Of course, in such artificial construction there are no
paradoxes with the evolution of the pure state into a mixed one.

The Minkowski vacuum is a translation invariant state, therefore any
quantization scheme implying this vacuum as the initial state should not give
a physical preference to the last ray with respect to other locations at the
initial stages of the collapse. From this it follows that the condition of
translational invariance of the vacuum requires that any quantization be
unitarily equivalent to the one which uses modes which are smooth and regular
on the last ray. However, it is well known that in the outer space such modes
are going to die away exponentially as the Schwarzschild time tends to
infinity. Hence it is impossible to have a late time stationary radiation flux
in these modes.

This means that the traditional scheme corresponds to some exotic system,
whose possibility for a physical realization is highly questionable. In any
case such a system has nothing to do with a black hole.

It is worth mentioning that the matching condition of the field along the
\textquotedblleft last ray - horizon" line (i.e., through the point after
which the last ray turns into the horizon) requires for the field to be zero
also along the horizon if it is zero along the last ray. This indicates again
that such a construction has nothing in common with a black hole since the
horizon, as well as the last ray, consists of regular space-time points which
are free of any physical obstacles and there are no reasons to apply to these
points any boundary conditions. From this point of view the calculations of
the authors of the article \cite{RW} (together with all their followers) which
led them to the assertion about the existence of a late time flux are not
surprising since all these papers are based on the groundless assumption of
the zero boundary condition for the currents and energy-momentum tensor of the
field on the horizon. The surprise is that they are trying to interpret this
artifact as corresponding to a physical black hole.\footnote{Moreover, any
discussion of the boundary condition on the horizon in the course of the
approach proposed in \cite{RW} can make sense only after we agree that the
gravitational anomaly in the 2-dimensional space-time indeed exists. Above we
assumed that this is the case. However, even this is an open question
\cite{Nak}.}

One is led to the following question: if a physical Schwarzschild black hole
cannot create particles, how can we reconcile this with thermodynamics? The
answer is evident: a body which can only absorb and not radiate cannot reach
thermal equilibrium with its environment and so, from this viewpoint,
equilibrium thermodynamics does not apply to black holes. Consequently, no
physically sensible notions of thermodynamical temperature and entropy can be
defined for such an object. This is natural because at the semiclassical
level, the black hole has no intrinsic degrees of freedom to which the
statistical approach can be applied.

Beyond the semiclassical approach, however, quantum gravity effects may start
to play a role and a black hole may acquire quantum degrees of freedom.
Nevertheless, even for such a case, as is shown rigorously by Y. Leblanc and
B. Harms \cite{Leb},\cite{Ha1},\cite{Ha2} similar conclusions still hold. The
thermal partition function associated with traditionally known
\textquotedblleft black hole thermodynamics" in fact does not exist. It is
indeed infinite for the entire temperature range and the interpretation of the
black hole instanton period as the inverse canonical temperature is wrong.
Then the whole \textquotedblleft black hole thermodynamics" collapses,
together with the concept of black hole entropy. Instead, in accordance with
the Leblanc-Harms theory, a quantum black hole looks like the p-branes
resonant excitations at the Planck scale \textit{without a classical horizon
structure}. The point is that this represents a \textit{pure state quantum
theory} and no statistical meaning whatsoever is attached to it. As a
consequence, there is no trace of any \textquotedblleft information loss
paradox" in this theory.

\section{Acknowledgements}

I would like to thank R. Jantzen for valuable comments and improvement of
English, V. Mur for discussion on \textquotedblleft time-like
tunneling\textquotedblright\ and G. Vereshchagin for the computer design of
the figures.

\end{document}